\documentclass[aip,
 amsmath,amssymb,
 reprint,
]{revtex4-1}

\usepackage{graphicx}
\usepackage{dcolumn}
\usepackage{bm}

\usepackage[utf8]{inputenc}
\usepackage[T1]{fontenc}
\usepackage{mathptmx}

\begin{document}

\preprint{AIP/123-QED}

\title[Imaging non-standing spin-waves in confined micron-sized ferromagnetic structures under uniform excitation]{Imaging non-standing spin-waves in confined micron-sized ferromagnetic structures under uniform excitation}

\author{S. Pile}
\affiliation{Institute of Semiconductor and Solid State Physics, Johannes Kepler University Linz, 4040 Linz, Austria}

\author{T. Feggeler}
\affiliation{Faculty of Physics and Center for Nanointegration Duisburg-Essen (CENIDE), University of Duisburg-Essen, 47057 Duisburg, Germany}

\author{T. Schaffers}
 \altaffiliation[Current address: ]{NanoSpin, Department of Applied Physics, Aalto University School of Science, P.O. Box 15100, FI-00076 Aalto, Finland.}
\affiliation{Institute of Semiconductor and Solid State Physics, Johannes Kepler University Linz, 4040 Linz, Austria}

\author{R. Meckenstock}
\affiliation{Faculty of Physics and Center for Nanointegration Duisburg-Essen (CENIDE), University of Duisburg-Essen, 47057 Duisburg, Germany}

\author{M. Buchner}
\affiliation{Institute of Semiconductor and Solid State Physics, Johannes Kepler University Linz, 4040 Linz, Austria}

\author{D. Spoddig}
\author{B. Zingsem}
 \altaffiliation[Also at ]{3Ernst Ruska-Centrum für Mikroskopie und Spektroskopie mit Elektronen, Forschungszentrum Jülich GmbH, D-52425 Jülich, Germany.}
\affiliation{Faculty of Physics and Center for Nanointegration Duisburg-Essen (CENIDE), University of Duisburg-Essen, 47057 Duisburg, Germany}

\author{V. Ney}
\affiliation{Institute of Semiconductor and Solid State Physics, Johannes Kepler University Linz, 4040 Linz, Austria}

\author{M. Farle}
\author{H. Wende}
\affiliation{Faculty of Physics and Center for Nanointegration Duisburg-Essen (CENIDE), University of Duisburg-Essen, 47057 Duisburg, Germany}

\author{H. Ohldag}
\affiliation{Stanford Synchrotron Radiation Laboratory, SLAC National Accelerator Laboratory, Menlo Park, California 94025, USA}

\author{A. Ney}
\affiliation{Institute of Semiconductor and Solid State Physics, Johannes Kepler University Linz, 4040 Linz, Austria}

\author{K. Ollefs}
\affiliation{Faculty of Physics and Center for Nanointegration Duisburg-Essen (CENIDE), University of Duisburg-Essen, 47057 Duisburg, Germany}

\date{\today}

\begin{abstract}
A non-standing character of directly imaged spin-waves in confined micron-sized ultrathin permalloy (Ni\textsubscript{80}Fe\textsubscript{20}) structures is reported along with evidence of the possibility to alter the observed state by modifications to the sample geometry. Using micromagnetic simulations the presence of the spin-wave modes excited in the permalloy stripes along with the quasi-uniform modes were calculated. The predicted spin-waves were imaged in direct space using time resolved scanning transmission X-ray microscopy, combined with a ferromagnetic resonance excitation scheme (STXM-FMR). STXM-FMR measurements revealed a non-standing character of the spin-waves. Also it was shown by micromagnetic simulations and confirmed with STXM-FMR results that the observed character of the spin-waves can be influenced by the local magnetic fields in different sample geometries.
\end{abstract}

\maketitle

The substitution of electrons by photons or quasi-particles such as magnons is of a great importance for potential alternative developments in future computing technologies \cite{001}. This technology will move much closer towards real applications, if one is able to actually make use of these quasi-particles. A prerequisite for that are well-investigated and understood properties of the particles, and thus the possibility to control their behaviour \cite{002,003,004}. The scope of our research is the investigation of magnetization dynamics as it provides an opportunity for entirely novel magnetic memory and logic devices \cite{005,006,007}.

Magnons or spin-waves being the elementary excitations of coupled spin systems in solids, have been widely investigated in thin films \cite{008,009,010}, nanostructured multilayers \cite{011}, magnonic crystals \cite{012,013,014,015} and magnonic waveguides \cite{016,017,018}. Spin-waves in confined micron-sized structures has also drawn increasing interest \cite{019,020} as understanding of their dynamics is utmost important for development of computational devices the size of which is only decreasing over the time.

The experimental investigation on such a small scale requires very sensitive experimental techniques. Moreover, if one wants to obtain information about time and space resolved magnetization dynamics, the measuring techniques are required to be quite sophisticated. The possibility to excite and visualize spin-waves in micron-sized structures was shown to be possible by techniques such as Brillouin light scattering (BLS) \cite{020,021} or spatially resolved ferromagnetic resonance force microscopy (FMRFM) \cite{022}. Later time resolved scanning transmission X-ray microscopy (TR-STXM) \cite{027,028,029} has been combined with a phase-locked ferromagnetic resonance (FMR) excitation scheme (STXM-FMR). This novel STXM-FMR technique enables direct time dependent imaging of the spatial distribution of the precessing magnetization over the sample during FMR excitation \cite{030, 031,032,033}.

In most of the cases spin-waves are investigated using a non-uniform excitation of the structure \cite{002, 004, 010, 012}. When the uniform excitation field is applied to the specimen it is also possible to excite spin-waves, but only standing spin-waves are expected \cite{034}. A standing spin-wave implies that its nodes nodes remain at the same position in space over the time. In case when the out of plane component of the in-plane precessing magnetization forms the wave, the nodes are the regions with zero out of plane magnetization. Development of planar micro-resonators allowed to measure the spin-wave spectrum of a single micron-sized stripe using conventional FMR technique \cite{023,024,025,026}. Using this technique the presence of spin-waves in a single micron-sized permalloy stripe during its uniform excitation was evidenced earlier with the support of micromagnetic simulations \cite{025}. To be detectable by FMR the spin-waves in the micron-sized stripe need to exhibit a standing character and thus provide a net absorption signal under excitation with a homogeneous microwave field. These findings were corroborated by detailed investigations of comparable spin-wave modes in a Co micron-sized stripe with similar dimensions \cite{026}.

For the STXM-FMR measurements to be able to directly image spin-wave FMR modes as reported previously \cite{025,026} we used a model system of two lithographically produced permalloy identical micron-sized stripes to investigate different orientations of the stripe relative to the external magnetic field under the same conditions. Each of the stripes has a lateral dimension of 1*5\,$\mu$m\textsuperscript{2} and a thickness of 30\,nm with 5\,nm Al capping to protect permalloy against oxidization. The stripes were oriented perpendicularly to each other with a gap of 2\,$\mu$m between them. In the chosen experimental geometry the external static magnetic field $B_{ext}$ was oriented in-plane along the length of one stripe ("easy orientation") and perpendicular to the length of the other ("hard orientation") \cite{026}. Two sample systems were taken into consideration: the so called "T-geometry" with the stripe in easy orientation centred to the middle of the length of the stripe in hard orientation (see Fig.\,2(b), Fig.\,3(a)), and "L-geometry" with the stripe in easy orientation aligned to one of the shorter edges of the stripe in hard orientation (see Fig.\,3(c)).

\begin{figure}
\includegraphics[width=0.47\textwidth]{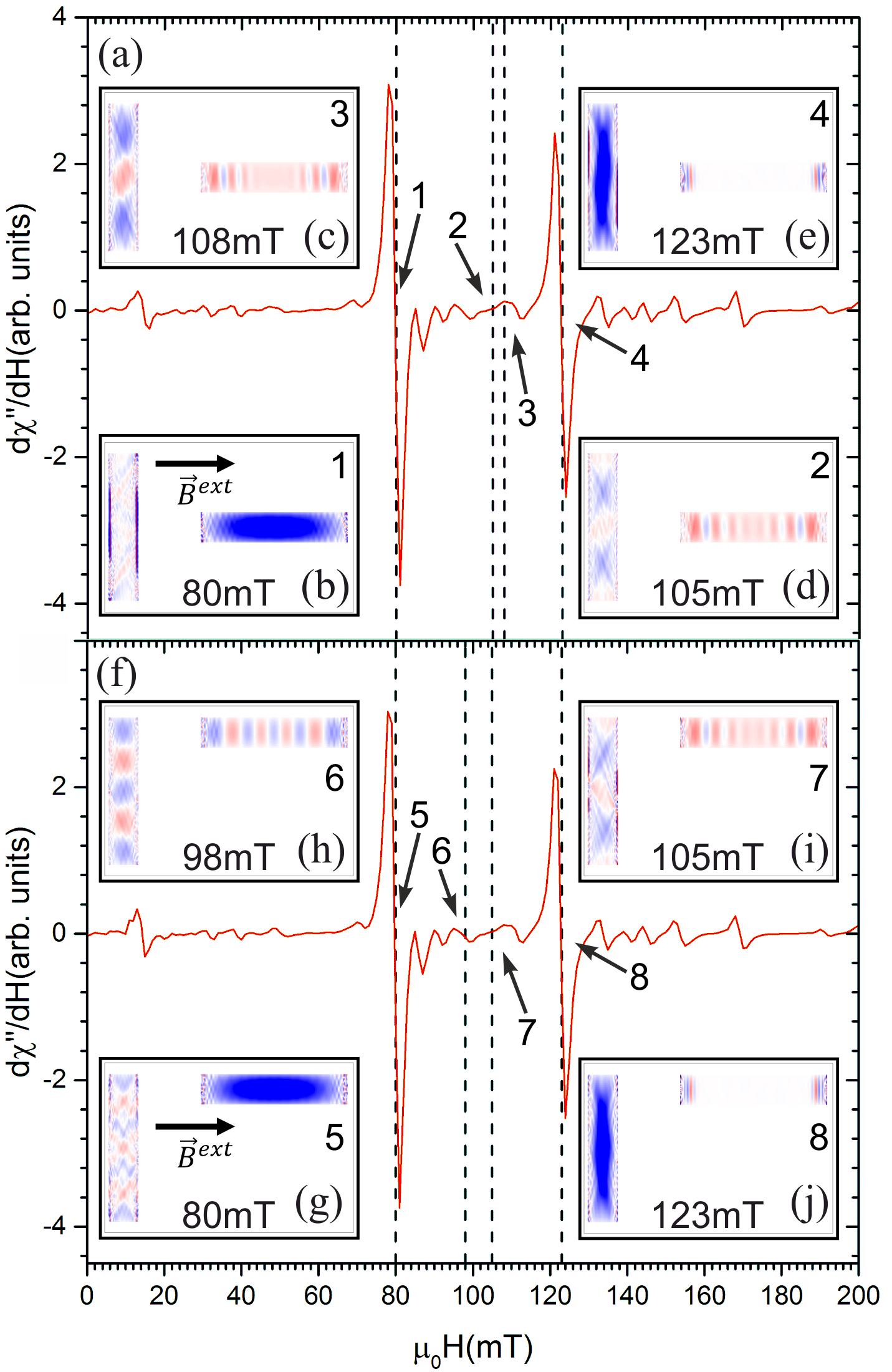}
\caption{\label{fig:epsart} MuMax3 simulations of FMR spectrum and the spatial distribution of the out-of-plane component of the magnetization in the permalloy micron-sized stripes in (a)-(d) T-geometry and (f)-(j) L-geometry.}
\end{figure}

To determine $B_{ext}$ at which the FMR quasi-uniform and spin-wave modes occur at the microwave frequency of $f_{MW}=9.446$\,GHz the MuMax3 \cite{035} micromagnetic simulations of the FMR spectrum was calculated for both geometries of the samples. In Fig.\,1(a) and (f) simulated FMR spectra of the T- and L-geometries are shown, respectively. Additionally the spatial distribution of the magnetic excitations was calculated at each simulated point of the spectra. The direction of the external field relative to the samples' orientation is shown in the insets (b) and (g) of the figure. Comparison of the simulated FMR spectra of both sample systems shows the same resonance field positions (see Fig.\,1(a) and (f)). Simulated spatial distributions of the magnetic excitations are displayed in the insets (c)-(d) and (h)-(j) in Fig.\,1. Red and blue colour in the images indicate opposite orientation of the dynamic out of plane component of the magnetization, the value of the component is indicated by the colour intensity. The white colour in the simulation images represent zero out of plane magnetization component. The spatial distribution in Fig.\,1 is shown for several external magnetic field values. The results of the simulations show that the biggest FMR signals at 80\,mT (signals no. 1 and 5) and 123\,mT (signals no. 4 and 8) are quasi-uniform modes of the stripes in easy and hard orientation, respectively. The separation in the field values for the quasi-uniform modes of two perpendicular stripes of one sample is a result of a shape anisotropy. The smaller signals observed in the spectra are the spin-wave modes of the two stripes \cite{025,026}.

Although there was no difference in the FMR spectra the spatial distribution of the magnetic excitations shows distinct difference between T- and L-geometries at some external field values. One of the examples is shown in the insets (d) and (i) in Fig.\,1 for the external field of 105\,mT. The spatial distribution of the magnetic excitations in the stripe in hard orientation shows asymmetry of the nodes along the length of the stripe in the L-geometry in comparison to the T-geometry, where the nodes are situated symmetrically along the stripe length with respect to the stripe center.

FMR modes are excited by applying a small periodic magnetic field perpendicular to the direction of the static field $B_{ext}$ which is applied in the film plane \cite{023,024,025,026}. When the FMR condition is fulfilled, a persistent precession of the magnetization around the direction of the effective magnetic field occurs. In case of in-plane magnetization (see Fig.\,2(b)) the dynamic component of the precessing magnetization perpendicular to the sample surface can be probed in the STXM-FMR by circularly polarized X-ray photons of the certain energy passing through the sample perpendicular to its plane as shown schematically in Fig.\,2(a) \cite{030,031}. The transmitted light intensity depends on the relative directions of the X-rays’ polarization and dynamic component of the magnetization, the so-called X-ray magnetic circular dichroism (XMCD) effect. Thus, the maximum and minimum intensities on the STXM-FMR images (white and black colours in the scans) mean maximum deviation of the transmitted X-rays and hence dynamic component of the magnetization in opposite directions.

The STXM-FMR measurements were carried out at a Ni L\textsubscript{3}-edge at the microwave excitation frequency of $f_{MW}=9.446$\,GHz. An example of STXM-FMR measurement of the stripe in hard orientation (marked with red rectangle in Fig.\,2(b)) in T-geometry sample is demonstrated in upper row of the images (STXM-FMR scans) in Fig.\,2(c). The scans represent 6 equidistant time-steps of the magnetization precession cycle depicting the dynamics of the quasi-uniform mode at the 123\,mT. In the plots the 7th point in time is always included which is a repetition of the results of the first one to demonstrate the full period of the recorded signal. Each scan includes stripe area marked with a red rectangle, and a background (outside the red rectangle). Under each STXM image in Fig.\,2(c) a snapshot of the corresponding micromagnetic simulation is shown, which resembles the time-evolution of excitation no. 4 in Fig.\,1(a) and (e). One can directly observe that at the quasi-uniform mode the magnetic moments of the sample precess in phase across almost its entire area. That can be seen from the STXM-FMR images where the magnetic contrast across the sample area changes its colour from white (0\,ps in Fig.\,2(c)) to black (52.9\,ps in Fig.\,2(c)) after a half of the precession period of 105.9\,ps while the background contrast remains the same. The non-uniformity of the precession closer to the stripe edges is due to a non-uniform effective field inside the sample originating from the strong demagnetizing fields closer to the stripe's edges \citep{019}. The overall agreement between the dynamic magnetic contrast of the STXM-FMR measurement and the micromagnetic simulation is remarkable and can be taken as experimental proof of the validity and credibility of the simulations.

\begin{figure}
\includegraphics[width=0.4\textwidth]{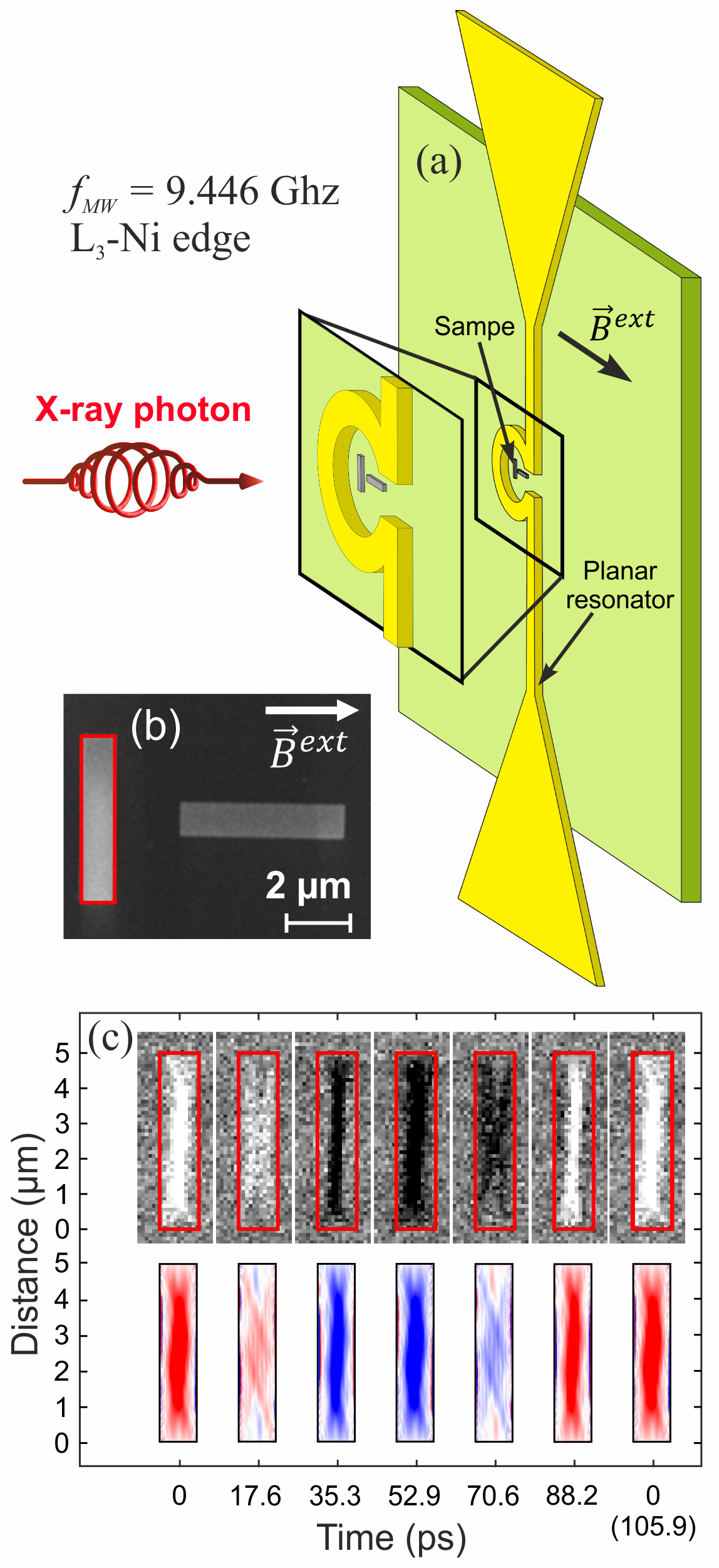}
\caption{\label{fig:wide} (a) Sketch of the STXM-FMR experiment geometry, (b) scanning electron microscope image of the sample in T-geometry, (c) results of the STXM-FMR measurements at 123\,mT for the stripe in hard orientation in the T-geometry sample, and corresponding micromagnetic simulations.}
\end{figure}

The STXM-FMR results for the spin-wave mode of the T-geometry sample at 108\,mT are shown in Fig.\,3(b). One can clearly distinguish three regions within the stripe (inside the red rectangle) with opposite orientation of the dynamic magnetization component. This indicates the presence of a spin-wave mode at the given external field. The red dots on the scans are used as a guide to the position of the nodes of the observed spin-wave. It is visible in the plot that the nodes do not remain at the same position within the stripe over the time which indicates that no purely standing spin-wave is observed. The same behaviour can be seen in the simulated spatial distribution plotted below in Fig.\,3(b) following the white spaces between red and blue regions in the stripe. An external field of 108\,mT corresponds to the signal no. 3 in Fig.\,1(a). At this field the sample is not at its FMR mode but 3\,mT below the closest spin-wave mode at 111\,mT. Hence, spin excitation observed in the STXM-FMR measurement is not a single eigenmode\cite{020}. Instead, inhomogeneities produced by internal magnetization landscapes of the stripe and inhomogeneous external magnetic stray fields, caused by the second stripe, can lead to a superposition of spin-wave eigenmodes resulting in the observed movement of the nodes. The observed movement is an evidence of the possibility to excite non-standing spin-waves in confined structures using uniform excitation (meaning uniform microwave field). Moreover further measurements showed that it is possible to alter the movement by modifying the mutual positions of the stripes moving from T- to L-geometry.

\begin{figure*}
\includegraphics[width=0.8\textwidth]{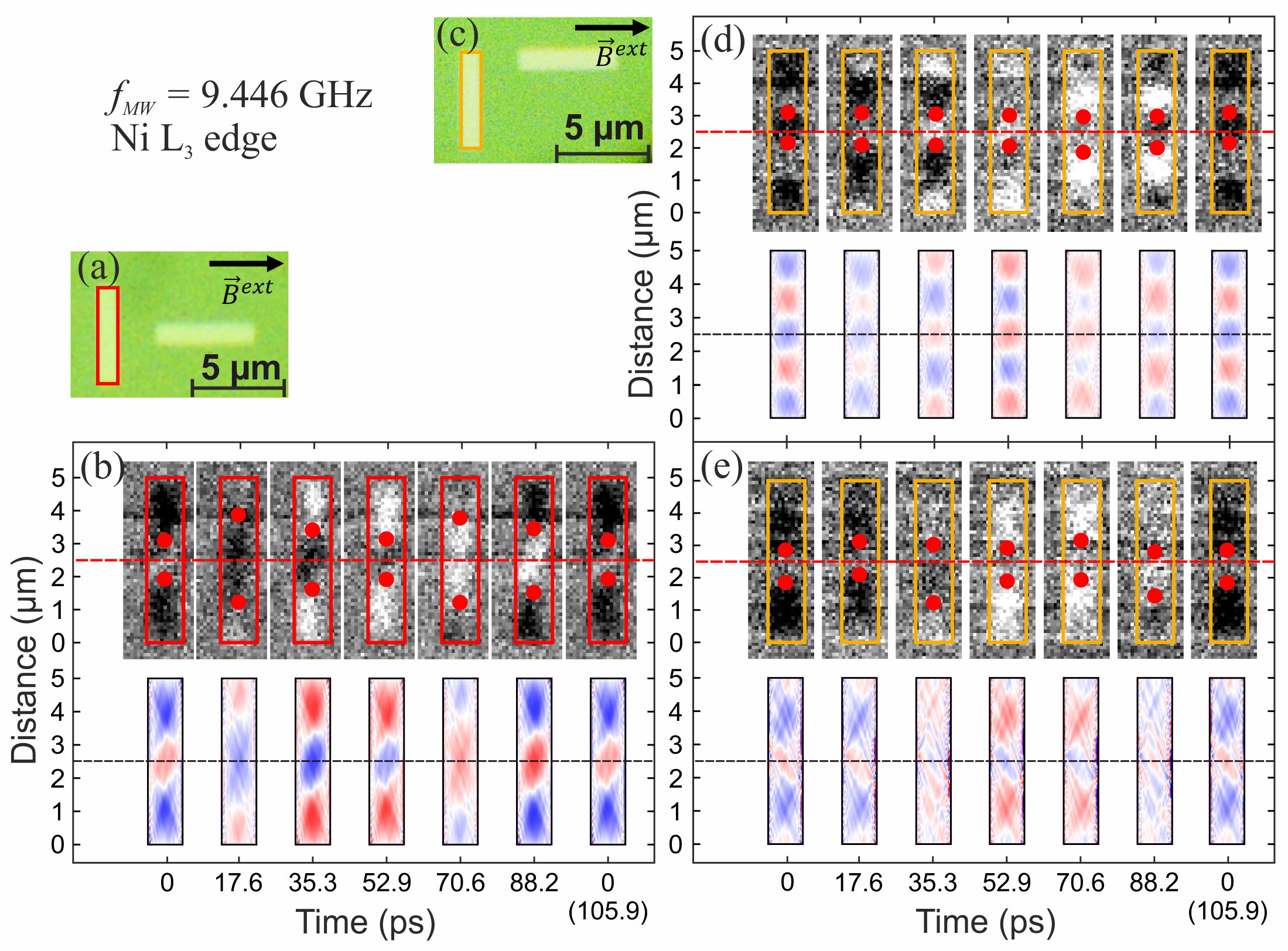}
\caption{\label{fig:wide}(a) Optical image of the sample in T-geometry, (b) results of the STXM-FMR measurements at 108\,mT for the stripe in hard orientation, and corresponding micromagnetic simulations, (c) optical image of the sample in L-geometry, (b) results of the STXM-FMR measurements at 98\,mT and (e) 105\,mT for the stripe in hard orientation, and corresponding micromagnetic simulations.}
\end{figure*}

The STXM-FMR measurements of the L-geometry sample were carried out at the same frequency and X-rays' energy as for the T-geometry. The results of two of these measurements are shown in Fig.\,3(d) and (e) at the fields of 98\,mT and 105\,mT, respectively. The stripe area is marked with orange rectangles in the STXM-FMR images. As one can see the mode at 98\,mT (see Fig.\,3(d)) is very close to a standing one as the position of the nodes marked with red dots does not change significantly along the stripe over the time. This observation is confirmed by simulations plotted below the STXM-FMR scans in Fig.\,3(d). Also the FMR simulation in Fig.\,1(f) (signal no. 6) show that signal no. 6 is only 1\,mT off from the spin-wave FMR mode at 97\,mT.

Results of the second measurement at 105\,mT exhibited non-standing character of the observed spin-wave, which can be concluded from the change of the position of the nodes in Fig.\,3(e). Notably the observed movement of the nodes at this field in L-geometry is asymmetric relative to the center of the stripe marked with the red dashed horizontal line in Fig.\,3(e). That can be seen, for example, when comparing scans at 0\,ps and 17.6\,ps points in time in Fig.\,3(e): from the first scan to the second both nodes are shifted slightly upwards. The field of 105\,mT corresponds to the signal no. 7 in Fig.\,1(f) which is 6\,mT below the closest spin-wave FMR mode. As discussed before a superposition of the spin-wave eigenmodes takes place in combination with the inhomogeneity of the external stray fields being shifted with the second stripe to the upper shorter edge of the discussed stripe. Therefore, the asymmetric behaviour of the spin-wave dynamics can be linked to the inhomogeneity of the external field which in case of the L-geometry is asymmetric.

Directly observed spin-waves are the result of the interference of the waves in several directions \cite{002,008,036}. It was reported earlier that edge roughness of the sample, edge surface profile \cite{018}, its shape \cite{002} or micro variations of the external field applied during excitation \cite{037} can cause significant changes in the spin-wave dispersion in one or several directions leading to different behaviour of the resulting wave. In case of the mode no. 6 in Fig.\,3(f) we still observe a slight variation of the nodes‘ position in space, which can be a result of not perfectly hitting the resonance field value and/or a roughness of the sample edges. It is possible that with a very small shift of the resonance field the standing character of the wave changes to the non-standing one when the transition and/or superposition between the eigenmodes occurs. This kind of movement is not possible to detect using conventional FMR technique but possible to image with STXM-FMR as was demonstrated above. Additional influence of the second stripe can lead to local disturbances, affecting the spin-wave dispersion in one or several directions and, hence, this leads to an observable asymmetry of the nodes' movement of the resulting wave.

In conclusion, our STXM-FMR experiments have shown that it is possible to directly observe quasi-uniform and spin-wave FMR modes in micron-sized permalloy stripes, as well as spin-waves in between the FMR modes. More importantly, we demonstrated the possibility to remotely alter the spin-wave behaviour in the stripe in the hard orientation, by placing the second stripe (easy orientation) at different positions along its length. Sample's geometry manipulation allows to modify magnetic field inhomogeneity within one of the stripes, thus leading to the change of the spin-wave behaviour. The work has a great potential as permalloy is a very widely used material and a shape modification is much easier achievable than the modification of the crystalline structure required for a number of materials used for inducing propagating spin-waves.

\begin{acknowledgements}
Use of the Stanford Synchrotron Radiation Lightsource, SLAC National Accelerator Laboratory, is supported by the U.S. Department of Energy, Office of Science, Office of Basic Energy Sciences under Contract No. DE-AC02-76SF00515. The authors would like to thank the Austrian Science Foundation (FWF), project No I-3050 as well as the German Research Foundation (DFG), project No OL513/1-1 for financial support.
\end{acknowledgements}

\nocite{*}
\bibliography{slReferences}% Produces the bibliography via BibTeX.

\end{document}